\documentclass[onecolumn, journal,11pt]{IEEEtran}

\IEEEoverridecommandlockouts
\title{New Results on Secret Key Establishment over a Pair of Broadcast Channels}
\author{Hadi Ahmadi, Reihaneh Safavi-Naini \\
\footnotesize{Department of Computer Science, University of Calgary, Canada.}\\
\footnotesize{\{hahmadi, rei\}@ucalgary.ca}}
\usepackage{multirow}
\usepackage{color}
\usepackage{cite}
\usepackage{epsfig}
\usepackage{graphicx}
\graphicspath{{../figures/}}
\usepackage{subfigure}
\usepackage{dsfont}
\usepackage{amssymb}
\usepackage{amsmath}
\usepackage{enumerate}

\newcommand{\remove}[1]{}

\newcommand{\tab}{\hspace*{3em}}
\newtheorem{theorem}{Theorem}
\newtheorem{corollary}{Corollary}
\newtheorem{lemma}{Lemma}
\newtheorem{definition}{Definition}
\newtheorem{proposition}{Proposition}

\begin{document}
\maketitle

\renewcommand{\baselinestretch}{1.2}
\normalsize
\begin{abstract}
The problem of Secret Key Establishment (SKE) over a pair of independent Discrete Memoryless Broadcast Channels (DMBCs) has already been studied in \cite{Ah10}, where we provided lower and upper bounds on the secret-key capacity. In this paper, we study the above setup under each of the following two cases: (1) the DMBCs have secrecy potential, and (2) the DMBCs are stochastically degraded with independent channels. In the former case, we propose a simple SKE protocol based on a novel technique, called Interactive Channel Coding (ICC), and prove that it achieves the lower bound. In the latter case, we give a simplified expression for the lower bound and prove a single-letter capacity formula under the condition that one of the legitimate parties can only send i.i.d. variables.
\end{abstract}

\section{Introduction}
We consider the following problem of Secret Key Establishment (SKE): Alice and Bob want to share a secret key in the presence of an eavesdropping adversary, Eve. Information-theoretic solutions to this problem assume that a collection of sources and/or channels are available to the parties. We refer this as a \emph{setup}.

Wyner's pioneering work \cite{Wy75} and its generalization by Csisz$\mathrm{\acute{a}}$r and K$\mathrm{\ddot{o}}$rner \cite{Cs78} considered transmission of secure messages over a Discrete Memoryless Broadcast Channel (DMBC) from Alice to Bob and Eve. They defined the secrecy capacity in this setup as the highest rate of secure and reliable message transmission (in bits per channel use) and showed that this capacity is positive if Bob's channel is less noisy \cite{Ko77} than Eve's. The work in \cite{Wy75,Cs78} has also been proved for the case of Gaussian channels \cite{Le78}. These results can also be used for SKE since any secure message transmission protocol can be used to send a secret-key securely over the DMBC.

Extensions of the work in \cite{Wy75,Cs78} have investigated the improvement of SKE by considering new setups. Maurer \cite{Ma93} and independently Ahlswede and Csisz$\mathrm{\acute{a}}$r \cite{Ah93} studied SKE when there is a DMBC from Alice to Bob and Eve, and a public discussion channel between Alice and Bob that is reliable, insecure, and unlimitedly available in both directions. They also considered SKE when the DMBC above is replaced by a Discrete Memoryless Multiple Source (DMMS) between the parties. Csisz$\mathrm{\acute{a}}$r and Narayan \cite{Cs00} considered SKE in the latter setup with a slight difference that the public channel is one-way and limited in rate. Ahlswede and Cai \cite{Ah06} studied SKE when Wyner's setup is accompanied by an additional secure (and reliable) \emph{output feedback channel} that is used to feed back the information received from the forward channel. Noisy feedback over modulo-additive broadcast channels is another extension \cite{Te08, La08}. Khisti et al. \cite{Kh08} and independently Prabhakaran et al. \cite{Pr08} considered a setup where the parties have access to a DMMS and a DMBC from Alice to Bob and Eve.

In practice special types of channel, e.g., public discussion channel, must be realized from more basic resources such as a DMBC. In \cite{Ah10}, we introduced a new setup for SKE, called \emph{2DMBC}, where the only resources available to Alice and Bob are two independent DMBCs in the two directions. This setup is appropriate to model wireless networks where two nodes can communicate interactively and their communication is eavesdropped by their wireless neighbors. The secret-key capacity in this setup is defined as the maximum rate of secure and reliable key establishment, in bits per channel use. Lower and upper bounds on the secret-key capacity in the 2DMBC setup have been provided and shown to coincide when the broadcast channels are \emph{physically degraded} \cite{Ah10}.

\subsection{Our work}
Motivated by applying the theoretical results to practical communication scenarios, in this paper, we extend the results of \cite{Ah10} in the following directions.

1) We consider the 2DMBC setup when both DMBCs have \emph{secrecy potential}, by which, we mean that realizing a noiseless channel from any of the DMBCs is not optimal. In most of the channels of interest (in communication), this occurs when the DMBCs have non-zero secrecy capacities. We propose a two-round SKE protocol based on a novel technique, called \emph{Interactive Channel Coding (ICC)} that achieves the lower bound in \cite{Ah10}. This lower bound was proved before by a SKE protocol that, although being convenient for the proof, uses an elaborate two-level coding construction whose efficient design becomes a new challenge in practice. Instead, ICC is a simple extension of systematic channel coding to a two-round construction in which the messages are essentially a codeword from a systematic error correcting code, split into two parts: one received in the first round and one sent in the second round. Roughly speaking, the ICC protocol works as follows. Alice sends a random sequence $R_A$ and Bob receives a noisy version of it, $I_A$. He chooses an independent random sequence, $I_B$, and appends it to $I_A$. We refer to the concatenated sequence $I=(I_A||I_B)$ as the \emph{information sequence}. Bob uses his systematic encoder to calculate a \emph{parity-check sequence} $P$ for the information sequence $I$, and sends $(I_B||P)$ to Alice, where Alice receives $(R_B||R_P)$. She uses her systematic decoder to decode $R=(R_A||R_B||R_P)$ to $\hat{I}=(\hat{I}_A||\hat{I}_B)$ as an estimation of the information sequence. The rest is to generate a secure key from the information sequence. ICC is particularly important as it allows progress in systematic capacity achieving codes to be directly applied to SKE.

2) We study the 2DMBC setup when the DMBCs are \emph{stochastically degraded with independent channels}. We refer to this setup as \emph{sd-2DMBC}. This study is motivated by observing that the results in \cite{Ah10} for the secret-key capacity of (physically) degraded 2DMBCs do not necessarily hold for stochastically degraded 2DMBCs. In setups like \cite{Pr08,Kh08,Cs00,Cs78} that do not offer interactive communication, physically and stochastically degraded broadcast channels are equivalent in terms of the secret-key capacity. This is not true, however, for the 2DMBC setup in which interactive communication is permitted. Two important classes of stochastically degraded channels with independent components are binary symmetric broadcast channels and Gaussian broadcast channels. We note that our results can be easily extended to continuous memoryless channels.
\begin{enumerate}[2-a)]
      \item We give a simplified expression for the lower bound on the secret-key capacity in the sd-2DMBC setup which uses fewer random variables and hence results in a simpler maximization problem.
      \item We consider sd-2DMBC when one of the parties can only send only independently, identically distributed (i.i.d) variables. We prove a single-letter formula for the secret-key capacity that is achieved by a two-round protocol.
\end{enumerate}
An example of the scenario (2-b) is when a base station wants to establish keys with several users in different locations. The offline computation power of the base station is high but its realtime computation power is limited. So, the base station sends i.i.d. variables in realtime and stores the received variables from all other nodes in all communication rounds. Next, it calculates the common keys with each user from the stored information in the offline mode. Our study of the above scenario provides a solution to this problem.

\subsection{Notation}
We use calligraphic letters $(\mathcal{U})$ to denote finite alphabets (sets), and the corresponding letters in uppercase $(U)$ and lowercase $(u)$ to denote random variables (RVs) and their realizations, respectively. The size of $\mathcal{U}$ is denoted by $|\mathcal{U}|$. $\mathcal{U}^n$ is set of all sequences of length $n$ whose elements are in $\mathcal{U}$; $U^n=(U_1,U_2,\dots,U_n)$ is called an $n$-sequence, i.e., a sequence of $n$ (possibly correlated) RVs in $\mathcal{U}$, and $U_i^j$ is used to denote a part of this sequence that is $(U_i, U_{i+1}, \dots, U_j)$. We use `$||$' to show the concatenation of sequences. For a value $x$, we use $[x]_+$ to show $\max\{0,x\}$. For three random sequences $Q_1$, $Q_2$, and $Q_3$, we use $Q_1 \leftrightarrow Q_2 \leftrightarrow Q_3$ to denote a Markov chain between them in this order.

\smallskip
\subsection{Paper organization}
Section \ref{sec_definition} describes the 2DMBC setup, definitions, and existing SKE results in this setup. Section \ref{sec_main results} summarizes the main results of this paper. Section \ref{sec-proofs} is dedicated to the proofs. We conclude the paper in Section \ref{sec_conclusion}.

\section{Model, Definitions, and Existing Results}\label{sec_definition}
\begin{figure} [h]
\centering
  \includegraphics[scale=.4]{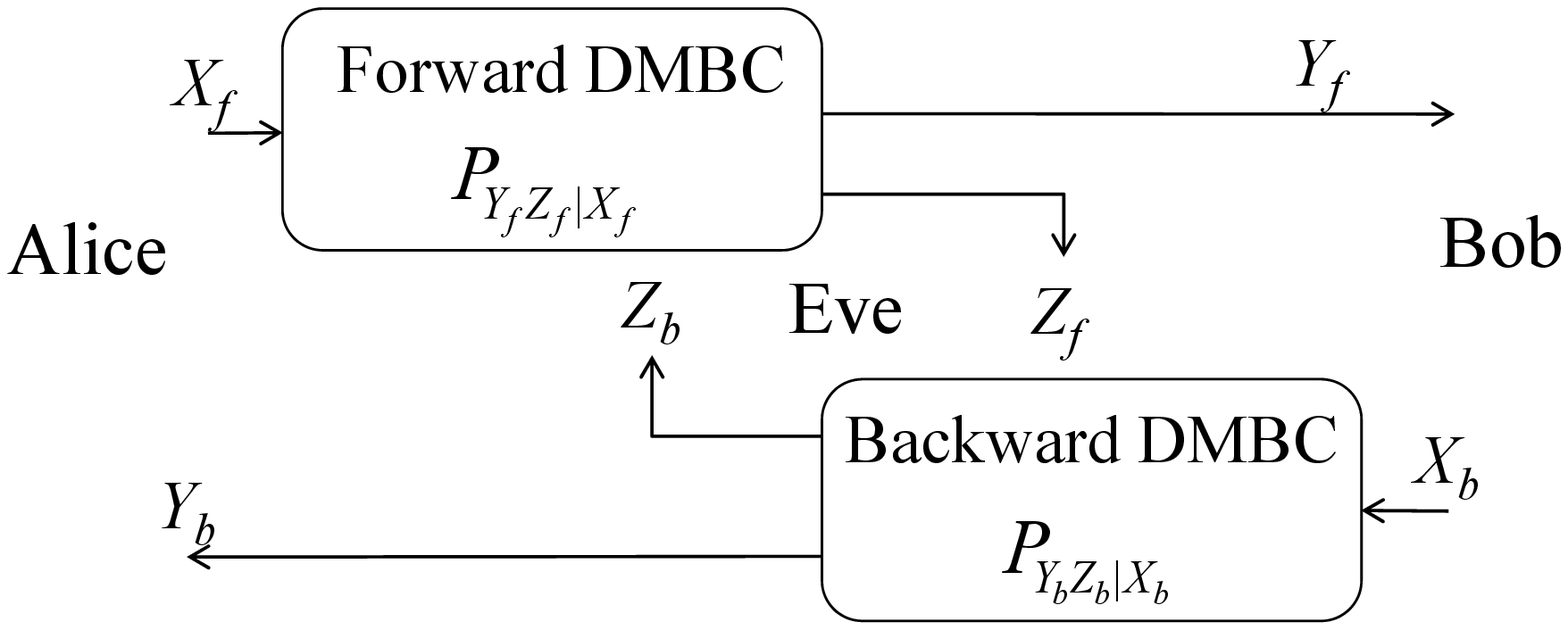}
  \caption{The 2DMBC setup}\label{fig_two-way}
\end{figure}

The 2DMBC setup is depicted in Fig. \ref{fig_two-way}. There is a forward DMBC, $X_f\rightarrow (Y_f,Z_f)$ specified by $P_{Y_f,Z_f|X_f}$, from Alice to Bob (and Eve) and a backward DMBC, $X_b\rightarrow (Y_b,Z_b)$ specified by $P_{Y_b,Z_b|X_b}$, from Bob to Alice (and Eve). We assume that each party has free access to an independent source of randomness.

An SKE protocol in this setup may contain several communication rounds. In each round either Alice or Bob sends a sequence of random variables (RVs) which is computed using some independent randomness and the communicated (sent and/or received) sequences in the previous rounds. Finally each party will have a set of communicated sequences, which form their \emph{view}. Using their views, one of the legitimate parties computes a key $S$, and the other one computes an estimation of the key $\hat{S}$. A secure SKE protocol and the secret-key capacity in the 2DMBC setup are defined as follows.
\begin{definition}\label{definition-secure SKE} \cite{Ah10}
An SKE protocol $\Pi$ in the 2DMBC setup is $(R_{sk}, \delta)$\emph{-secure} if it results in the key $S$ and its estimation $\hat{S}$ such that
\begin{IEEEeqnarray}{l} \label{SKE-Eqs}
\frac{H(S)}{n_f+n_b} >  R_{sk}-\delta,\IEEEyessubnumber \label{SKE-rand} \\
\Pr(\hat{S}\neq S) < \delta,\IEEEyessubnumber \label{SKE-rel} \\
\frac{H(S|View_E)}{H(S)}>1-\delta,\IEEEyessubnumber\label{SKE-sec}
\end{IEEEeqnarray}
where $View_E$ is Eve's view at the end of the protocol, and $n_f$ and $n_b$ are the number of times that the forward and the backward channels are used, respectively.
\end{definition}

\begin{definition}\label{definition-secrecy capacity}\cite{Ah10}
The \emph{secret-key capacity} in the 2DMBC setup, $C^{2DMBC}_{sk}$, is the largest $R_{sk} \geq 0$ such that, for any arbitrarily small $\delta>0$, there exists an $(R_{sk}, \delta)$-secure SKE protocol.
\end{definition}

We recall the lower and the upper bounds given in \cite{Ah10} on the secret-key capacity in the 2DMBC setup. Let the RVs $X_f,Y_f,Z_f$ (resp. $X_b,Y_b,Z_b$) correspond to the conditional distribution $P_{Y_f,Z_f|X_f}$ (resp. $P_{Y_b,Z_b|X_b}$), specified by the 2DMBC. Let $V_{f}$, $V_{b}$, $W_{1,f},W_{2,f}$, $W_{1,b},W_{2,b}$ be RVs from arbitrary sets where $V_{f}$, $V_{b}$, $ (W_{1,f},W_{2,f})$, and $(W_{1,b},W_{2,b})$ are independent and the following Markov chains are  satisfied:
\begin{IEEEeqnarray}{l} \label{Markov-Chain}
V_{f}\leftrightarrow Y_f \leftrightarrow (X_f,Z_f), ~
W_{2,b}\leftrightarrow W_{1,b}\leftrightarrow X_b \leftrightarrow (Y_b,Z_b),\tab \IEEEyessubnumber \label{Markov-Chain-1} \\
V_{b} \leftrightarrow Y_b \leftrightarrow (X_b,Z_b), ~
W_{2,f}\leftrightarrow W_{1,f}\leftrightarrow X_f \leftrightarrow (Y_f,Z_f). \tab \IEEEyessubnumber \label{Markov-Chain-2}
\end{IEEEeqnarray}
Also let
\begin{IEEEeqnarray}{l}\label{R^AB_{sk}12}
R^A_{s1}=I(V_{f};X_f)-I(V_{f};Z_f),\IEEEyessubnumber \label{RA_{sk}1} \\
R^A_{s2}=I(W_{1,b};Y_b|W_{2,b})-I(W_{1,b};Z_b|W_{2,b}),\IEEEyessubnumber \label{RA_{sk}2}\\
R^B_{s1}=I(V_{b};X_b)-I(V_{b};Z_f),\IEEEyessubnumber \label{RB_{sk}1} \\
R^B_{s2}=I(W_{1,f};Y_f|W_{2,f})-I(W_{1,f};Z_f|W_{2,f})\IEEEyessubnumber. \label{RB_{sk}2}
\end{IEEEeqnarray}
The secret-key capacity is lower bounded \cite{Ah10} as
\begin{eqnarray} \label{lower-bound}
C^{2DMBC}_{sk} \geq \max \{L_{A},L_{B}\},
\end{eqnarray}
where
\begin{IEEEeqnarray}{l}
L_{A} =  \max_{n_f,n_b,P_{X_f,V_{f}},P_{X_b, W_{2,b},W_{1,b}}} \left[  \frac{n_f R^A_{s1}+ n_b [R^A_{s2}]_+}{n_f+n_b} \mathrm{s.~t.}~~ n_f I(V_{f};Y_f|X_f) < n_b I(W_{1,b};Y_b) \right],  \label{L_A} \\
L_{B} =  \max_{n_f,n_b,P_{X_b,V_{b}}, P_{X_f,W_{2,f},W_{1,f}}} \left[  \frac{n_b R^B_{s1} + n_f [R^B_{s2}]_+}{n_f+n_b} \mathrm{s.~t.}~~ n_bI(V_{b};Y_b|X_b) < n_f I(W_{1,f};Y_f) \right], \label{L_B}
\end{IEEEeqnarray}
and it is upper bounded \cite{Ah10} as
\begin{eqnarray} \label{upper-bound}
C^{2DMBC}_{sk} \leq \max_{P_{X_f},P_{X_b}} \{ I(X_f;Y_f|Z_f) , I(X_b;Y_b|Z_b) \}.
\end{eqnarray}

\section{Statement of Main Results}\label{sec_main results}
\subsection{The interactive channel coding protocol}\label{sec-main-ICC}
The lower bound in (\ref{lower-bound}) has been obtained by an SKE protocol \cite{Ah10} that uses a complicated two-level coding construction whose efficient design becomes a challenge in practice. We introduce the interactive channel coding (ICC) technique which is used to design the so-called \emph{ICC protocol} for SKE. We show that when the DMBCs have secrecy potential, the ICC protocol can achieve the lower bound in (\ref{lower-bound}). ICC relies on the existence of capacity-achieving \emph{systematic channel codes}. Designing efficient constructions for systematic channel codes has been well studied, e.g., a large body of work on the design of capacity achieving channel codes follows on linear block codes which can be represented as systematic codes. This makes the design of an efficient ICC protocol for SKE as simple as the design of efficient coding for SKE over a (one-way) DMBC \cite{Cs78}.

\begin{definition} \label{systematic code}
A \emph{(bipartite) systematic channel code}, with encoding alphabets $(\mathcal{Y}_f,\mathcal{X}_b)$ and decoding alphabets $(\mathcal{X}_f,\mathcal{Y}_b)$, is specified by a pair of encoding/decoding functions $(Enc/Dec)$, where
\begin{itemize}
\item $Enc:\mathcal{Y}_f^{n_f} \times \mathcal{X}_b^{n_{b,i}} \rightarrow \mathcal{Y}_f^{n_f} \times \mathcal{X}_b^{n_b}$ deterministically maps $(y_f^{n_f}||x_b^{n_{b,i}})$ (as the information sequence) to the codeword $(y_f^{n_f} ||x_b^{n_b})$ such that $x_b^{n_b}=(x_b^{n_{b,i}}||x_b^{n_{b,p}})$ and $n_b=n_{b,i}+n_{b,p}$; we call $x_b^{n_{b,p}}$ the parity-check sequence.
\item $Dec:\mathcal{X}_f^{n_f} \times \mathcal{Y}_b^{n_b} \rightarrow \mathcal{Y}_f^{n_f}\times \mathcal{X}_b^{n_{b,i}}$ deterministically assigns a guess $(\hat{y}_f^{n_f}||\hat{x}_b^{n_{b,i}})$ to each input $(x_f^{n_f}||y_b^{n_b})$.
\end{itemize}
\end{definition}

\begin{figure}[h]
\centering
  \includegraphics[scale=.4]{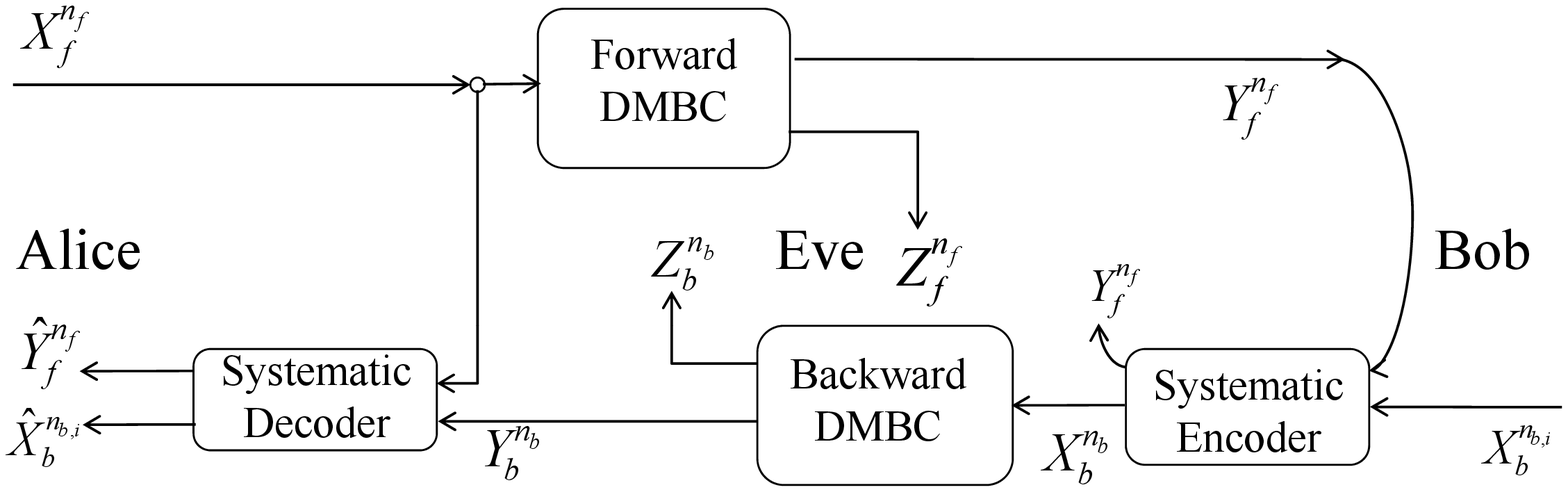}
  \caption{ICC over a 2DMBC: Alice initiates the protocol}\label{wits3}
\end{figure}

The general construction of the ICC protocol and a proof of Theorem \ref{theorem-ICC} are provided in Section \ref{sec-ICC-proof}. In the following, we describe the ICC protocol for a special case when $V_{f}=Y_f$, $W_{2,b}=1$, $W_{1,b}=X_b$, and Alice is the initiator (see Fig. \ref{wits3}). Accordingly, we rephrase the argument to be maximized and the constraint condition in (\ref{R^ICC-A}) respectively as
\begin{IEEEeqnarray}{c}
R_{sk}=\frac{n_f [I(Y_f;X_f)-I(Y_f;Z_f)] + n_b [I(X_b;Y_b)-I(X_b;Z_b)]}{n_f+n_b}, \label{R_sk-def}\\
n_f (H(Y_f|X_f)+\alpha)  \leq n_b I(X_b;Y_b) \label{const-cond-reph},
\end{IEEEeqnarray}
where $\alpha>0$ is an arbitrarily small constant. Let $n_b=n_{b,i}+n_{b,p}$, where $n_{b,i}$ is chosen to satisfy
\begin{IEEEeqnarray}{l} \label{n_b,i H}
n_{b,i} H(X_b) = n_b I(X_b;Y_b) - n_f (H(Y_f|X_f)+\alpha).
\end{IEEEeqnarray}
Let $N=n_f+n_b$ and $\epsilon$ be a small constant such that $5N\epsilon < n_f \alpha$. Let $\mathcal{Y}^{n_f}_{f,\epsilon}$ (resp. $\mathcal{X}^{n_{b,i}}_{b,\epsilon}$) be the set of all $\epsilon$-typical sequences w.r.t. $P_{Y_f}$ (resp. $P_{X_b}$) in $\mathcal{Y}_f^{n_f}$ (resp.  $\mathcal{X}_b^{n_{b,i}}$); Define
\begin{IEEEeqnarray*}{lll} \label{n_b,i H}
\eta_f = \log|\mathcal{Y}^{n_f}_{f,\epsilon}|,& \tab \eta_b = \log|\mathcal{X}^{n_{b,i}}_{b,\epsilon}|, \\
\eta=\eta_f+\eta_b,& \tab \kappa = N R_{sk}, & \tab \gamma = \eta - \kappa.
\end{IEEEeqnarray*}

Let $\{\mathcal{G}_i\}_{i=1}^{2^\kappa}$ be a partition of $\mathcal{Y}^{n_f}_{f,\epsilon} \times \mathcal{X}^{n_{b,i}}_{b,\epsilon}$ into $2^\kappa$ parts, each of size $2^\gamma$. Define $g:\mathcal{Y}^{n_f}_{f,\epsilon} \times \mathcal{X}^{n_{b,i}}_{b,\epsilon} \rightarrow \{1,2,\dots,2^\kappa\}$ as a function that, for every input $(y_f^{n_f},x_b^{n_{b,i}}) \in \mathcal{G}_i$, outputs $i$.

\noindent \textbf{\emph{Encoding.}} Alice chooses an i.i.d. $n_f$-vector $X_f^{n_f}$ and sends it over the forward DMBC; Bob and Eve receive $Y_f^{n_f}$ and $Z_f^{n_f}$, respectively. If $Y_f^{n_f} \notin \mathcal{Y}^{n_f}_{f,\epsilon}$, Bob returns a NULL; otherwise, he chooses uniformly at random an $n_{b,i}$-sequence $X_b^{n_{b,i}}$ from $\mathcal{X}^{n_{b,i}}_{b,\epsilon}$, encodes $Enc(Y_f^{n_f}||X_b^{n_{b,i}})=(Y_f^{n_f}||X_b^{n_b})$, and sends $X_b^{n_b}$ over the backward DMBC; Alice and Eve receive $Y_b^{n_b}$ and $Z_b^{n_b}$, respectively.

\noindent \textbf{\emph{Decoding.}} Alice decodes $(\hat{Y_f}^{n_f}||\hat{X}_b^{n_{b,i}})=Dec(X_f^{n_f}||Y_b^{n_b})$ using bipartite jointly typical decoding: she searches through the $2^{\eta}$ words in $\mathcal{Y}^{n_f}_{f,\epsilon}\times \mathcal{X}^{n_{b,i}}_{b,\epsilon}$ and either finds a unique $(\hat{Y}^{n_{f}}_f, \hat{X}_b^{n_{b,i}})$ such that $Enc(\hat{Y}^{n_{f}}_f, \hat{X}_b^{n_{b,i}})$ and $(X^{n_f}_f,Y^{n_b}_b)$ are $(n_f,\epsilon)$-bipartite jointly typical w.r.t. $(P_{Y_f,X_f},P_{X_b,Y_b})$ (see Section \ref{sec-ICC-proof}, Definition \ref{defbip_joi_typ}), or returns a NULL.

\noindent \textbf{\emph{Key derivation.}} Bob computes $S=g(Y_f^{n_f},X_b^{n_{b,i}})$. Alice computes $\hat{S}=g(\hat{Y_f}^{n_f},\hat{X}_b^{n_{b,i}})$.

\begin{theorem}\label{theorem-ICC}
Taking the variables from (\ref{Markov-Chain}) and (\ref{R^AB_{sk}12}), the ICC protocol can achieve the secret-key rate
\begin{eqnarray} \label{R^ICC}
R^{ICC}=\max \{R^{ICC}_{A}, R^{ICC}_{B} \},
\end{eqnarray}
where
\begin{IEEEeqnarray}{l}\label{R^ICC-A}
R^{ICC}_{A}=\max_{n_f,n_b,P_{X_f,V_{f}},P_{X_b, W_{2,b},W_{1,b}}} \{\frac{n_f R^A_{s1}+ n_b R^A_{s2}}{n_f+n_b} ~~\mathrm{s.~t.}~~ n_f[I(V_{f};Y_f|X_f)] < n_b I(W_{1,b};Y_b) \}, \tab\\
\nonumber \\
R^{ICC}_{B}=\max_{n_f,n_b,P_{X_b,V_{b}},P_{X_f, W_{2,f},W_{1,f}}} \{\frac{n_f R^B_{s1}+ n_b R^B_{s2}}{n_f+n_b} ~~\mathrm{s.~t.}~~ n_b[I(V_{b};Y_b|X_b)] < n_f I(W_{1,f};Y_f) \}. \tab \label{R^ICC-B}
\end{IEEEeqnarray}
\end{theorem}

Comparing (\ref{L_A}) with (\ref{R^ICC-A}), we conclude that $R^{ICC}_A$ and $L^A$ are equal if for the optimal selection of the parameters, in the maximization problem of (\ref{L_A}), $R^A_{s2}$ becomes non-negative. In other words, the two values (rates) are equal if the backward DMBC has secrecy potential, i.e., the optimal strategy is not based on realizing a noiseless channel from the backward DMBC. Similarly, $R^{ICC}_B$ equals $L^B$ if the forward DMBC has secrecy potential.
\begin{corollary}\label{final}
When the DMBCs have secrecy potential, the ICC protocol can achieve the lower bound in (\ref{lower-bound}).
\end{corollary}

\subsection{The secret-key capacity in the sd-2DMBC setup}
SKE over \emph{physically degraded} 2DMBCs (pd-2DMBCs) was considered in \cite{Ah10}, where we showed that the lower and the upper bounds coincide and the capacity is achieved by a one-round SKE protocol. This implies that interaction over a pd-2DMBC cannot increase the SKE rate. However, this is not generally true for \emph{stochastically degraded} broadcast channels, and the upper bound in (\ref{upper-bound}) does not necessarily coincide with the lower bound in (\ref{lower-bound}) for stochastically degraded DMBCs. In this paper,
we consider SKE over a 2DMBC, where each DMBC is \emph{stochastically degraded with independent channels}. We refer to this setup as \emph{sd-2DMBC}.
\begin{definition}\label{definition-stoch-degraded}
The DMBC $X\rightarrow (Y,Z)$, with conditional distribution $P_{YZ|X}$, is \emph{stochastically degraded in favor of $Y$} (or the party who receives $Y$) if there exist two RVs $\tilde{Y}$ and $\tilde{Z}$ such that $X\leftrightarrow \tilde{Y} \leftrightarrow \tilde{Z}$ forms a Markov chain and
\begin{IEEEeqnarray*}{l}
P_{XY}(x,y)=P_{X,\tilde{Y}}(x,y),\tab P_{XZ}(x,z)=P_{X,\tilde{Z}}(x,z).
\end{IEEEeqnarray*}
It consists of \emph{independent channels} if $P_{YZ|X}=P_{Y|X}.P_{Z|X}$.
\end{definition}
\begin{definition}\label{definition-sd-2DMBC}
A \emph{sd-2DMBC} is a 2DMBC whose DMBCs are stochastically degraded (either in favor of $Y$ or in favor of $Z$), and consist of independent channels.
\end{definition}

\subsubsection{Lower bound}
\begin{proposition} \label{proposition-lowerbound}
The secret-key capacity in the sd-2DMBC setup is lower bounded as
\begin{eqnarray} \label{SD-lower-bound}
C^{sd-2DMBC}_{sk} \geq \max \{L'_{A},L'_{B}\},
\end{eqnarray}
where
{\small
\begin{IEEEeqnarray}{ll}
L'_{A} &=  \max_{n_f,n_b,P_{V_f,X_f,X_b}} \{  \frac{n_f I(V_f;X_f|Z_f) + n_b [I(X_b;Y_b)-I(X_b;Z_b)]_+}{n_f+n_b}  ~\mathrm{s.~t.}~ n_f[I(V_{f};Y_f|X_f)] < n_b I(X_b;Y_b)\}, ~~~~~ \label{L'_A} \\
\nonumber \\
L'_{B} &=  \max_{n_f,n_b,P_{V_b,X_b,X_f}} \{  \frac{n_b I(V_b;X_b|Z_b) + n_f [I(X_f;Y_f)-I(X_f;Z_f)]_+}{n_f+n_b} ~\mathrm{s.~t.}~ n_b[I(V_{b};Y_b|X_b)] < n_f I(X_f;Y_f)\}. \label{L'_B}
\end{IEEEeqnarray}
}
\end{proposition}

The expressions (\ref{L'_A}) and (\ref{L'_B}) do not contain the RVs $W_{1,b},W_{2,b}, W_{1,f}$, and $W_{2,f}$, compared to (\ref{L_A}) and (\ref{L_B}). So, the maximization problem in obtaining the lower bound (\ref{SD-lower-bound}) is easier than that in (\ref{lower-bound}).

\subsubsection{single-letter characterization}
We consider a scenario where one of the legitimate parties can only send i.i.d. variables, and derive an expression for the secret-key capacity under this condition.
\begin{theorem} \label{theorem-SD-2DMBC}
When one of the legitimate parties can only send i.i.d. variables, the secret-key capacity in the sd-2DMBC setup equals
\begin{eqnarray} \label{C^sd-2DMBC}
\max \{L'_{A},L'_{B}\},
\end{eqnarray}
where $L'_{A}$ and $L'_{B}$ are given in (\ref{L'_A}) and (\ref{L'_B}), respectively.
\end{theorem}

\section{Proofs}\label{sec-proofs}
\subsection{Proof of Theorem \ref{theorem-ICC}, the ICC protocol}\label{sec-ICC-proof}
We describe the ICC protocol when Alice is the initiator and prove that it achieves the rate in (\ref{R^ICC-A}). In a similar way, one can describe ICC when Bob initiates the protocol and prove (\ref{R^ICC-B}). First we give the following definitions from \cite{Ah10} for \emph{bipartite typical sequences}. A bipartite sequence $X^N=(U^n||T^d)$, where $N=n+d$, is the concatenation of two subsequences, $U^n\in \mathcal{U}^n$ and $T^d \in \mathcal{T}^d$, with two probability distributions, $P_{U^n}$ and $P_{T^d}$, respectively.

\begin{definition}
A sequence $x^N=(u^n||t^d)$ is an \emph{$(\epsilon, n)$-bipartite typical sequence} with respect to the probability distribution
pair $(P_U(u), P_T(t))$, iff
\begin{eqnarray} \label{def-x-typical}
|-\frac{1}{N}\log P(x^N)-\frac{nH(U)+dH(T)}{N}|<\epsilon,
\end{eqnarray}
where $P(x^N)$ is calculated as
\begin{eqnarray}
\displaystyle P(x^N)=\prod_{i=1}^n
P_U(u_i)\times \prod_{i=1}^d P_T(t_i).
\end{eqnarray}
\end{definition}
\medskip

\begin{definition} \label{defbip_joi_typ}
A pair of sequences $(x^N,y^N)=((u^n||t^d),(u'^n||t'^d))$ is an
\emph{$(\epsilon, n)$-bipartite jointly typical pair of sequences} with
respect to the probability distribution pair $(P_{U,U'}(u,u'),
P_{T,T'}(t,t'))$, iff $x^N$ and $y^N$ are $(\epsilon, n)$-bipartite
typical sequences with respect to the marginal probability
distribution pairs $(P_{U}(u), P_T(t))$ and $(P_{U'}(u'),
P_T'(t'))$, respectively, and
\begin{eqnarray}\label{def-xy-typical}
|-\frac{1}{N}\log P(x^N,y^N)-\frac{nH(U,U')+dH(T,T')}{N}|<\epsilon,
\end{eqnarray}
where $P(x^N,y^N)$ is calculated as
\begin{eqnarray}
\displaystyle P(x^N,y^N)=\prod_{i=1}^n P_{U,U'}(u_i,u'_i)\times \prod_{i=1}^d P_{T,T'}(t_i,t'_i).
\end{eqnarray}
\end{definition}

Back to the proof, let the RVs $V_{f}, X_f, Y_f, Z_f$, and $W_{1,b}, W_{2,b}, X_b, Y_b, Z_b$ be the same as defined in Theorem \ref{theorem-ICC} such that the Markov chains in (\ref{Markov-Chain}) are satisfied. Also let $n_f$ and $n_b$ be integers that satisfy the constraint condition in (\ref{R^ICC-A}). For simplicity, we use $W_{1}, W_{2}$, and $V$ to refer to $W_{1,b}, W_{2,b}$, and $V_{f}$, respectively. Accordingly, we write the argument to be maximized in (\ref{R^ICC-A}) as
\begin{eqnarray}\label{R_s}
R_{sk}=\frac{n_f R^A_{s1} + n_b R^A_{s2} }{n_f+n_b}
\end{eqnarray}
where
\begin{IEEEeqnarray}{l}
R^A_{s1}=I(V;X_f)-I(V;Z_f), \IEEEyessubnumber \label{R^A_s1} \\
R^A_{s2}=I(W_{1};Y_b|W_{2})-I(W_{1};Z_b|W_{2}),\IEEEyessubnumber \label{R^A_s2}
\end{IEEEeqnarray}
and we rephrase the constraint condition in (\ref{R^ICC-A}) as
\begin{eqnarray}
n_b I(W_{1};Y_b) \geq n_f (I(V;Y_f|X_f)+ 3\alpha), \label{n_b-n_f}
\end{eqnarray}
where $\alpha>0$ is an small constant to be determined (later) from $\delta$. We shall show that for any given $\delta > 0$, for sufficiently large $n_f$ and $n_b$ that satisfy (\ref{n_b-n_f}), the three requirements in (\ref{SKE-Eqs}) can be satisfied.

Let $N=n_f+n_b$ and $\epsilon, \beta >0$ be small constants determined from $\alpha$ such that $3 N \epsilon < n_b \beta = n_f \alpha$. Let $n_b=n_{b,1}+n_{b,2}$, where $n_{b,2}$ is chosen to satisfy
\begin{equation} \label{n_{b,2}}
    n_{b,2} I(W_{1};Y_b) =  n_f (I(V;Y_f|X_f)+ 3\alpha).
\end{equation}
Define
\begin{IEEEeqnarray}{lll}
\eta_f=n_f [I(V;Y_f)+\alpha],& \eta_{f,2}=n_{b,2} I(W_2;Y_b),& \eta_{f,1}=\eta_f-\eta_{f,2}, \label{eta_f-def}\\
\eta_b=n_{b,1} [I(W_{1};Y_b) - \beta], \tab&  \eta_{b,2}=n_{b,1} I(W_{2};Y_b), \tab& \eta_{b,1}=\eta_b-\eta_{b,2}, \label{eta_b-def}\\
\eta_1=\eta_{f,1} + \eta_{b,1},& \eta_2=\eta_{f,2} + \eta_{b,2},& \eta=\eta_f + \eta_b, \label{eta_1-def}\\
\kappa=(n_f+n_b) R_{sk}, & \gamma=\eta-\kappa. \label{kappa&gamma-def}
\end{IEEEeqnarray}
Although the quantities obtained in (\ref{n_{b,2}})-(\ref{kappa&gamma-def}) are real values, for sufficiently large $n_b$ and $n_f$, we can approximate them by integers. Since $\beta$ can be made arbitrarily small, we can assume $\eta_b$ and $\eta_f$ are non-negative. Furthermore, since
\begin{IEEEeqnarray*}{ll}
\eta  = \eta_f + \eta_b
        & \stackrel{(a)}{=} n_f [I(V;Y_f,X_f)+\alpha] + n_{b,1} [I(W_1,Y_b) - \beta] \\
        & = n_f I(V;X_f)+ n_f I(V;Y_f|X_f)+ n_f \alpha + n_{b,1} I(W_1,Y_b) - n_{b,1} \beta \\
        & \stackrel{(b)}{=} n_f I(V;X_f) + n_{b,2} I(W_1,Y_b) - 2 n_f \alpha + n_{b,1} I(W_1,Y_b) - n_{b,1} \beta \\
        & \geq n_f I(V;X_f) + n_{b} I(W_1,Y_b) - 3 n_f \alpha \geq R^A_{s1} + R^A_{s2} - 3 n_f \alpha \\
        & \geq \kappa- 3 n_f \alpha,
\end{IEEEeqnarray*}
for arbitrarily small $\alpha$, we can assume $\eta\geq \kappa$ and so $\gamma$ is non-negative. Equality (a) above is due to (\ref{eta_f-def}), (\ref{eta_b-def}), and the Markov chain $X_f\leftrightarrow Y_f \leftrightarrow V$, and equality (b) follows from (\ref{n_{b,2}}). The following sets and functions are used in the design of the ICC protocol.
\begin{enumerate}[(i)]
    \item $\mathcal{V}^{n_f}$ is the set of all possible $n_f$-sequences with elements from $\mathcal{V}$. Create $\mathcal{V}_{\epsilon}^{n_f}$ by randomly and independently selecting $2^{\eta_f}$ $\epsilon$-typical sequences (w.r.t. $P_V$) from $\mathcal{V}^{n_f}$.
    \item Let $\mathfrak{f}:\mathcal{V}_{\epsilon}^{n_f} \rightarrow \mathcal{F}=\{1,2,\dots,2^{\eta_f}\}$ be an arbitrary bijective mapping; denote its inverse by $\mathfrak{f}^{-1}$.
    \item let $\{\mathcal{F}_i\}_{i=1}^{2^{\eta_{f,2}}}$ be a partition of $\mathcal{F}$, into $2^{\eta_{f,2}}$ equal-sized parts. Label elements of part $i$ as $\mathcal{F}_i=\{f_{i,j}\}_{j=1}^{\eta_{f,1}}$. Define $\mathfrak{f_{ind}}:\mathcal{F}\rightarrow \{1,\dots,2^{\eta_{f,2}}\} \times \{1,\dots,2^{\eta_{f,1}}\}$ such that $\mathfrak{f_{ind}}(f)=(i,j)$, if $f$ is labeled by $f_{i,j}$.
    \item $\mathcal{W}_1^{n_{b,1}}$ is the set of all possible sequences $W_1^{n_{b,1}}$. Create $\mathcal{W}_{1,\epsilon}^{n_{b,1}}$ by randomly selecting $2^{\eta_b}$ different $\epsilon$-typical sequences (w.r.t. $P_{W_1}$) from $\mathcal{W}_1^{n_{b,1}}$.
    \item Let $\mathfrak{b}:\mathcal{W}_{1,\epsilon}^{n_{b,1}} \rightarrow \mathcal{B}=\{1,2,\dots,2^{\eta_b}\}$ be an arbitrary bijective mapping; denote its inverse by $\mathfrak{b}^{-1}$.
    \item In analogy to $\mathcal{F}$, let $\{\mathcal{B}_i\}_{i=1}^{2^{\eta_{b,2}}}$ be a partition of $\mathcal{B}$ where $\mathcal{B}_i=\{b_{i,j}\}_{j=1}^{2^{\eta_{b,1}}}$. Define $\mathfrak{b}_{indx}:\mathcal{B}\rightarrow \{1,\dots,2^{\eta_{b,2}}\} \times \{1,\dots,2^{\eta_{b,1}}\}$ such that $\mathfrak{b}_{indx}(b)=(i,j)$, if $b$ is labeled by $b_{i,j}$.
    \item Let $\{\mathcal{G}_i\}_{i=1}^{2^\kappa}$ be a partition of $\mathcal{F} \times \mathcal{B}$ into parts of size $2^{\gamma}$. Define $g:\mathcal{F} \times \mathcal{B} \rightarrow \{1,2,\dots,2^\kappa\}$ such that, for any input in $\mathcal{G}_i$, it outputs $i$.
    \item Define the parity-check book $\mathcal{P}_{2}$ as a the collection of $2^{\eta_2}$ words $\{w^{n_{b,2}}_{2,f_2,b_2}:~f_2=1,2,\dots,2^{\eta_{f,2}}, ~b_2=1,2,\dots,2^{\eta_{b,2}}\}$, where each codeword $w^{n_{b,2}}_{2,f_2,b_2}$ is of length $n_{b,2}$ and is independently generated according to the distribution
         \[\prod_{i=1}^{n_{b,2}} p(W_2=w_{2,f_2,b_2}(i)).\]
    \item For each $w^{n_{b,2}}_{2,f_2,b_2}$, Define the parity-check book $\mathcal{P}_1(w^{n_{b,2}}_{2,f_2,b_2})$ as a the collection of $2^{\eta_1}$ words $\{w^{n_{b,2}}_{1,f_2,b_2,f_1,b_1}\\ : ~f_1=1,\dots,2^{\eta_{f,1}}, ~b_1=1,\dots,2^{\eta_{b,1}}\}$, where each codeword $w^{n_{b,2}}_{1,f_2,b_2,f_1,b_1}$ is of length $n_{b,2}$ and is independently generated according to the distribution
        \[ \prod_{i=1}^{n_{b,2}} p(W_1=w_{1,f_2,b_2,f_1,b_1}(i)|W_2=w_{2,f_2,b_2}(i)). \]
    \item Let $Enc:\mathcal{V}^{n_f}\times \mathcal{W}_1^{n_{b,1}} \rightarrow \mathcal{V}^{n_f} \times \mathcal{W}_1^{n_b}$ be a (bipartite) systematic encoding function such that $Enc(v^{n_f},w_1^{n_{b,1}})=(v^{n_f},w_1^{n_b})$, where $w_1^{n_b} = (w_1^{n_{b,1}},w^{n_{b,2}}_{1,f_2,b_2,f_1,b_1})$, using the above parity-check books when $f=\mathfrak{f}(v^{n_f})$, $b=\mathfrak{b}(\mathcal{W}_1^{n_{b,1}})$, $(f_2,f_1)=\mathfrak{f_{ind}}(f)$, and $(b_2,b_1)=\mathfrak{b_{ind}}(b)$.
    \item Let $DMC_W$ be the DMC, $W_1\rightarrow X_b$, that is specified by $P_{X_b|W_1}$.
\end{enumerate}

\bigskip

\noindent \textbf{Encoding.} Alice selects an i.i.d. $n_f$-sequence $X^{n_f}_f$ and sends it over the forward DMBC. Bob and Eve receive $Y_f^{n_f}$ and $Z_f^{n_f}$, respectively. Bob finds a $V^{n_f} \in \mathcal{V}_{\epsilon}^{n_f}$ that is $\epsilon$-jointly typical with $Y_f^{n_f}$ (w.r.t. $P_{V,Y_f}$), or returns a NULL if he fails. He selects independently a uniformly random $W_1^{n_{b,1}} \in \mathcal{W}_{1,\epsilon}^{n_{b,1}}$. He computes $F=\mathfrak{f}(V^{n_f})$, $B=\mathfrak{b}(W_1^{n_{b,1}})$, $(F_2,F_1)=\mathfrak{f_{ind}}(F)$, and $(B_2,B_1)=\mathfrak{b_{ind}}(B)$, and calculates $Enc(V^{n_f},W_1^{n_{b,1}})=(V^{n_f},W_1^{n_b})$ using these variables. Next, Bob inputs $W_1^{n_{b}}$ to $DMC_W$ to compute $X^{n_{b}}_b$, and sends $X^{n_{b}}_b$ over the backward DMBC. Alice and Eve receive $Y_b^{n_{b}}$ and $Z^{n_{b}}_b$, respectively.
\medskip

\noindent \textbf{Decoding.} Alice searches through $\mathcal{V}^{n_f}_{\epsilon}\times \mathcal{W}^{n_{b,1}}_{1,\epsilon}$ and either finds a unique $(\hat{V}^{n_f},\hat{W}_1^{n_{b,1}})$ that is $(\epsilon, {n_f})$-bipartite jointly typical to $(X^{n_f}_f,Y^{n_b}_b)$ w.r.t. $(P_{V,X_f},P_{W_1,Y_b})$, or returns a NULL.
\medskip

\noindent \textbf{Key Derivation.} Bob computes $S=g(F,B)$. Alice computes $\hat{F}=\mathfrak{f}(\hat{V}^{n_f})$ and $\hat{B}=\mathfrak{b}(\hat{W}_1^{n_{b,1}})$, and then $\hat{S}=g(\hat{F},\hat{B})$.
\medskip

\begin{figure}[ht]
\centering
\subfigure[Encoding and decoding]{
\includegraphics[scale=.36]{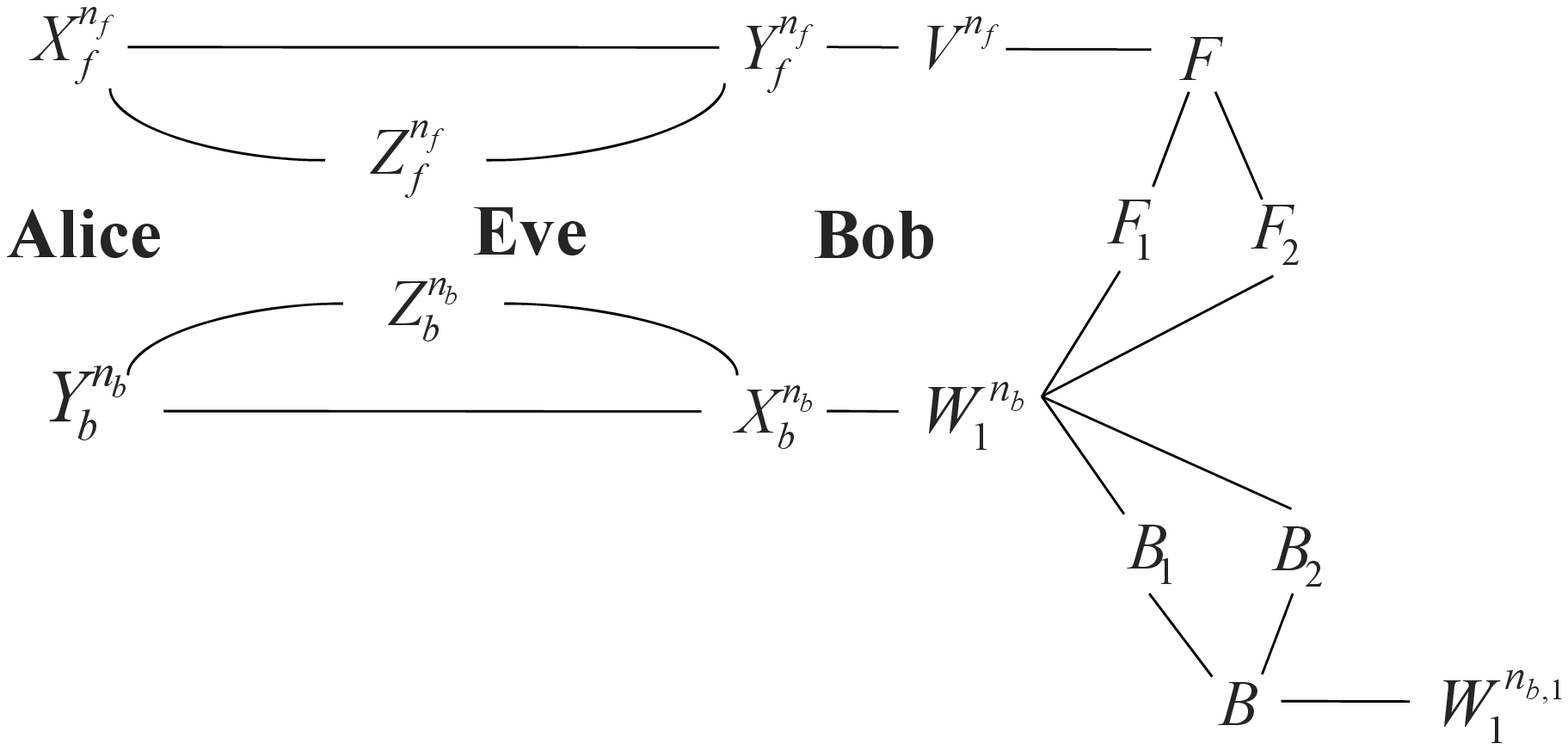}
\label{codec_ICC}
}
\subfigure[Key derivation by Alice]{
\includegraphics[scale=.36]{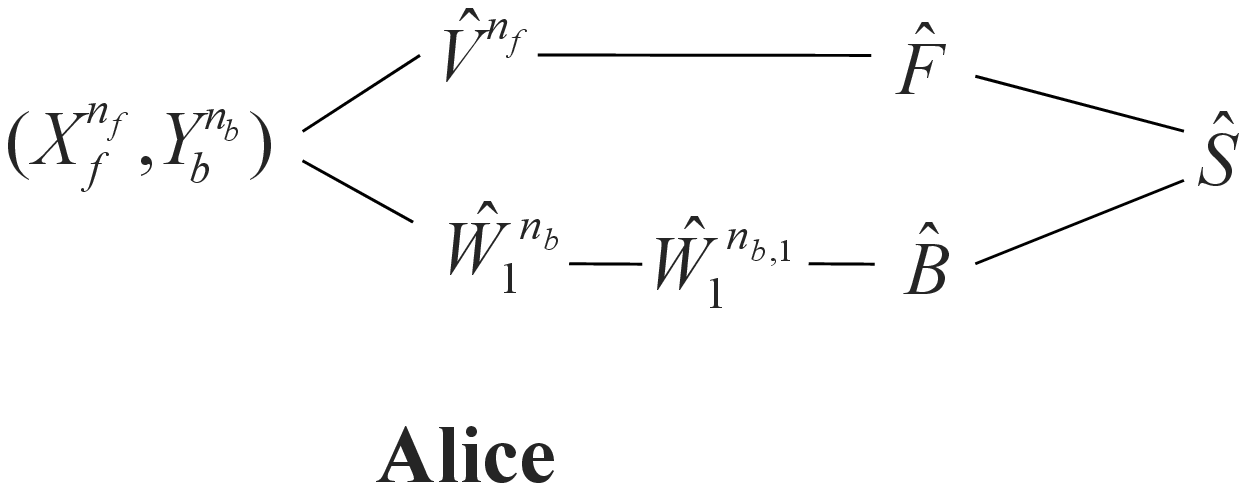}
\label{key_estimate_ICC}
}
\subfigure[Key derivation by Bob]{
\includegraphics[scale=.4]{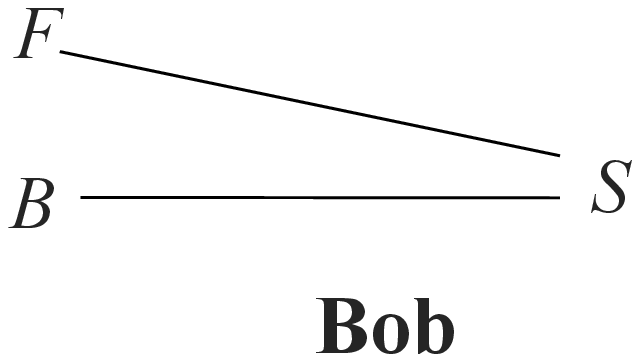}
\label{key_derivate_ICC}
}
\caption[Optional caption for list of figures]{The relation between the variables/sequences used in the ICC protocol for \subref{codec_ICC} encoding/decoding, \subref{key_estimate_ICC} key derivation by Alice, and \subref{key_derivate_ICC} key derivation by Bob}
\label{figSKE-ICC}
\end{figure}

Fig. \ref{figSKE-ICC} shows the relationship between the random variables/sequences used in the ICC protocol. Two variables/sequences are connected by an edge if (1) they belong to input/outputs of the same DMBC, or (2) one is computed from the other by Alice or Bob using a (possibly randomized) function.

\noindent \textbf{Uniformity Analysis: Proving (\ref{SKE-rand})}\\
From AEP for $P_V$ (see \cite[Appendix A]{Ah10} for more details), and since $F$ and $V^{n_f}$ have the same distribution,
\begin{IEEEeqnarray}{l}
\forall f \in \mathcal{F},~ \Pr(F=f) \leq 2^{-\eta_f + 5 N \epsilon}. \label{P_F}\\
\Rightarrow \eta_f - 5 N \epsilon \leq H(V^{n_f})=H(F) \leq \eta_f, \label{H_F}
\end{IEEEeqnarray}
Since $W_1^{n_{b,1}}$ (resp. $B$) is selected uniformly at random from $\mathcal{W}_{1,\epsilon}^{n_{b,1}}$ (resp. $B$) of size $\eta_b$
\begin{IEEEeqnarray}{l}
\forall b \in \mathcal{B},~ \Pr(B=b) =  2^{- \eta_b}\\
\Rightarrow H(W_1^{n_{b,1}})=H(B) = \eta_b. \label{H_B}
\end{IEEEeqnarray}

For every $i\in \{1,2,\dots,{2^\kappa}\}$, the probability that $S=i$ equals to the probability that $(F,B)\in \mathcal{G}_i$. More specifically (see (\ref{eta_1-def}) and (\ref{kappa&gamma-def})),
\begin{IEEEeqnarray}{l}
\forall i:~ \Pr(S=i) = \sum_{f,b \in \mathcal{G}_i} \Pr(F=f \wedge B=b) \leq 2^{\gamma} 2^{-\eta_f + 5 N \epsilon} 2^{-\eta_b} = 2^{\gamma} 2^{-\eta + 5 N \epsilon}
                     =2^{- (\kappa - 5 N \epsilon)}\nonumber\\
\Rightarrow \frac{H(S)}{n_f+n_b} \geq \frac{\kappa - 5 N \epsilon}{n_f+n_b} = R_{sk} - \delta,   \tab \delta \geq 5 \epsilon . \label{H_S}
\end{IEEEeqnarray}
\medskip

\noindent \textbf{Reliability Analysis: Proving (\ref{SKE-rel})}\\
Since there are $\eta_f= n_f [I(V;Y_f)+\alpha]$ sequences in $\mathcal{V}^{n_f}_{\epsilon}$, from joint-AEP, with probability arbitrarily close to 1, there exists a $V^{n_f} \in \mathcal{V}^{n_f}_{\epsilon}$ that is $\epsilon$-jointly typical with $Y_f^{n_f}$ (w.r.t. $P_{V,Y_f}$) and the encoding phase is successful. In the decoding phase, Alice needs to search through the $2^{\eta}$ words in $\mathcal{V}^{n_f}_{\epsilon}\times \mathcal{W}^{n_{b,1}}_{1,\epsilon}$, where $\eta$ is calculated as
\begin{eqnarray} \label{eta}
\eta    &=& \eta_f + \eta_b \stackrel{(a)}{=} n_f (I(V;Y_f)+\alpha) + n_{b,1} (I(W_{1};Y_b) - \beta) \nonumber \\
        &\stackrel{(b)}{=}& n_f (I(V;Y_f)+\alpha) + n_b I(W_{1};Y_b) - n_f (I(V_{f};Y_f|X_f)+ 3\alpha) - n_{b,1} \beta \nonumber \\
        &\stackrel{(c)}{=}& n_f(I(V;X_f,Y_f)+\alpha) + n_b I(W_{1};Y_b) - n_f (I(V_{f};Y_f|X_f)+ 3\alpha) - n_{b,1} \beta  \\
        &=& n_f I(V;X_f) + n_b I(W_{1};Y_b) - 2 n_f \alpha - n_{b,1} \beta \nonumber \\
        & <& n_f I(V;X_f) + n_b I(W_1;Y_b) - 9 N \epsilon. \label{eta_2}
\end{eqnarray}
Equality (a) follows from (\ref{eta_f-def}) and (\ref{eta_b-def}), equality (b) follows from (\ref{n_{b,2}}), and equality (c) is due to the Markov chain $X_f\leftrightarrow Y_f\leftrightarrow V$. Since $\eta$ is sufficiently smaller than $n_f I(V;X_f) + n_b I(W_1;Y_b)$, from AEP for bipartite sequences (see \cite[Theorem 4]{Ah10}), there exist an encoding function $Enc(.)$ for which the decoding error probability becomes arbitrarily close to 0. This implies that
\begin{eqnarray*}
\Pr(\hat{S}\neq S) \leq \Pr\left( (\hat{F},\hat{B}) \neq (F,B) \right) = \Pr\left( (\hat{V}^{n_f}, \hat{W}_1^{n_{b,1}}) \neq (V^{n_f},W_1^{n_{b,1}}) \right) < \delta.
\end{eqnarray*}
\medskip

\noindent \textbf{Secrecy Analysis: Proving (\ref{SKE-sec})}\\
We shall show that the $H(S|Z_f^{n_f},Z^{n_b}_b)$ is close to $H(S)$. For the quantities $H(F_2)$ and $H(B_2)$, we have (see \cite[Appendix A]{Ah10} for more details)
\begin{IEEEeqnarray}{l}
\eta_{f,2}- 5N \epsilon \leq H(F_2) \leq \eta_{f,2}, \label{H_F2} \\
\Rightarrow H(B_2) = \eta_{b,2}. \label{H_B2}
\end{IEEEeqnarray}

We write $H(S|Z_f^{n_f},Z^{n_b}_b)$ as
\begin{IEEEeqnarray}{lll}
H(S|Z_f^{n_f},Z^{n_b}_b)&\geq& H(S|F_2,B_2,Z_f^{n_f},Z^{n_b}_b) \nonumber \\
                        &=& H(S,F,B|F_2,B_2,Z_f^{n_f},Z^{n_b}_b)-H(F,B|S,F_2,B_2,Z_f^{n_f},Z^{n_b}_b) \nonumber \\
                        &=& H(F,B|F_2,B_2,Z_f^{n_f},Z^{n_b}_b)-H(F,B|S,F_2,B_2,Z_f^{n_f},Z^{n_b}_b)\nonumber \\
                        &=& H(F,B|F_2,B_2)-I(F,B;Z_f^{n_f},Z^{n_b}_b|F_2,B_2)-H(F,B|S,F_2,B_2,Z_f^{n_f},Z^{n_b}_b).~~  \label{HS-bound1}
\end{IEEEeqnarray}
The first term above is written as

The first term is written as
\begin{IEEEeqnarray}{l}
H(F,B|F_2,B_2)  = H(F|F_2,B_2) + H(B|F,F_2,B_2) \stackrel{(a)}{=} H(F|F_2) + H(B|B_2)  \nonumber \\
\tab \stackrel{(b)}{=} H(F) + H(B) -H(F_2) - H(B_2) \nonumber \\
\tab \stackrel{(c)}{\geq} \eta_f - 5N \epsilon + \eta_{b} - \eta_{F,2} - \eta_{b,2} \nonumber\\
\tab \stackrel{(d)}{\geq} n_f I(V;Y_f) - 2 N\epsilon + n_{b,1}[I(W_1;Y_b) - \beta]  - n_{b,2}I(W_2;Y_b) - n_{b,1}I(W_2;Y_b) \nonumber\\
\tab \stackrel{(e)}{=} n_f I(V;X_f) + n_f I(V;Y_f|X_f) -2 N \epsilon + n_{b,1} I(W_{1};Y_b) - n_{b}I(W_2;Y_b) - n_{b,1} \beta \nonumber \\
\tab = n_f I(V;X_f) + n_f (I(V;Y_f|X_f)+3\alpha) + n_{b,1} I(W_{1};Y_b) - n_{b}I(W_2;Y_b) - 3n_f \alpha - n_b \beta - 2 N \epsilon \nonumber \\
\tab \stackrel{(f)}{=} n_f I(V;X_f) + n_{b,2} I(W_{1};Y_b) + n_{b,1} I(W_{1};Y_b) - n_{b}I(W_2;Y_b) - 3n_f \alpha - n_b \beta - 2 N \epsilon\nonumber \\
\tab >  n_f I(V;X_f) + n_{b} I(W_{1};Y_b) - n_{b}I(W_2;Y_b) - 14 N \epsilon\nonumber \\
\tab \stackrel{(g)}{=} n_f I(V;X_f) + n_{b} I(W_{1};Y_b|W_2) - 14 N \epsilon \label{HS-bound2}
\end{IEEEeqnarray}
Equality (a) holds since $B_2$ and $B$ are selected independently of $F_2$ and $F$, equality (b) holds since $F_2$ and $B_2$ are deterministic functions of $F$ and $B$, respectively (the encoding phase), inequality (c) follows from (\ref{H_F}), (\ref{H_B}), (\ref{H_F2}), and (\ref{H_B2}), equality (d) follows from (\ref{eta_f-def}) and (\ref{eta_b-def}), equality (e) is due to the Markov chain $X_f\leftrightarrow Y_f \leftrightarrow V$, equality (f) follows from (\ref{n_{b,2}}), and equality (g) is due to the Markov chain $W_2\leftrightarrow W_1 \leftrightarrow Y_b$.

The second term in (\ref{HS-bound1}) is written as
\begin{IEEEeqnarray}{lll}
I(F,B;Z_f^{n_f},Z^{n_b}_b|F_2,B_2)&=& I(F,B;Z_f^{n_f}|F_2,B_2) + I(F,B;Z^{n_b}_b|Z_f^{n_f},F_2,B_2) \nonumber\\
& \stackrel{(a)}{=}& I(V^{n_f},B;Z_f^{n_f}|F_2,B_2) + I(F,B;Z^{n_b}_b|Z_f^{n_f},F_2,B_2) \nonumber\\
& \stackrel{(b)}{\leq}& I(V^{n_f};Z_f^{n_f}) + I(F,B;Z^{n_b}_b|F_2,B_2) \nonumber\\
& \stackrel{(c)}{=}&            I(V^{n_f};Z_f^{n_f}) + H(Z^{n_b}_b|F_2,B_2)- H(Z^{n_b}_b|F,B) \nonumber\\
& \stackrel{(d)}{\leq}& {n_f} I(V;Z_f)  + n_b[H(Z_b|W_2)- H(Z_b|W_1)] \nonumber \\
& \stackrel{(e)}{\leq}& {n_f} I(V;Z_f)  + n_bI(W_1;Y_b|W_2) \label{HS-bound3}
\end{IEEEeqnarray}
Inequality (a) holds because $V^{n_f}=\mathfrak{f}^{-1}(F)$ (the key derivation phase), equality (b) is due to the Markov chains $(F_2,B_2)\leftrightarrow (V^{n_f},B) \leftrightarrow Z_f^{n_f}$, $B\leftrightarrow V^{n_f} \leftrightarrow Z_f^{n_f}$ and $Z_f^{n_f} \leftrightarrow F\leftrightarrow Z_b^{n_b}$, equality (c) holds since $F_2$ and $B_2$ are deterministic functions of $F$ and $B$, equality (d) follows from AEP, and equality (e) is due to the Markov chain $W_2\leftrightarrow W_1 \leftrightarrow Z_b$.

It remains to calculate $H(F,B|S,F,B,Z_f^{n_f},Z_b^{n_b})$, i.e., the third term in (\ref{HS-bound1}). From (vii), knowing $S=i$ gives the partition $\mathcal{G}_{i}$ that $F,B$ belongs to; further, knowing $F_2=f_2$ and $B_2=b_2$ gives the parity-check sequence $w^{n_{b,1}}_{2,f_2,b_2} \in \mathcal{P}_2$ which is used in the encoding phase (see (viii)). Define the codebook
\[ \mathcal{C}^e_i = \{v^{n_f},w_1^{n_{b}}:~ (\mathfrak{f}(v^{n_f}),b) \in \mathcal{G}_i,~ w_1^{n_b}=Enc(\mathfrak{f}(v^{n_f}),b),~ F_2=f_2,~ B_2=b_2\}.\]
Given $S=i, Z_f^{n_f}$, and $Z^{n_b}_b$, one can search all the codewords in $\mathcal{C}^e_i$ and return a unique $\check{V}^{n_f},\check{W}_1^{n_{b}} \in \mathcal{C}^e_i$ that is $(\epsilon, {n_f})$-bipartite jointly typical to $(Z^{n_f}_f,Z^{n_b}_b)$ w.r.t. $(P_{V,Z_f},P_{W_1,Z_b})$; otherwise return a NULL. From (vii), $|\mathcal{G}_{i}|=2^\gamma$, and so $|\mathcal{C}^e_{i}|=2^{\gamma-\eta_2}$, where $\eta_2$ is given in (\ref{eta_1-def}). We first calculate $\eta$ which is used in the calculation of $\gamma-\eta_2$.

\begin{eqnarray*}
\eta    &=& \eta_f + \eta_b \\
        &=& n_f (I(V;Y_f) + \alpha) + n_{b,1} I(W_{1};Y_b) - n_b \beta \\
        &=& n_f I(V;X_f) + n_f (I(V;Y_f|X_f)+ 3\alpha) + n_{b,1} I(W_{1};Y_b) - 2 n_f \alpha - n_b \beta \\
        &=& n_f I(V;X_f) + n_b I(W_{1};Y_b) - 3 n_f \alpha.
\end{eqnarray*}
$\gamma-\eta_2$ is written as
\begin{eqnarray*}
\gamma-\eta_2 &\stackrel{(a)}{=}& \eta-({n_f}+{n_b})R_{sk} - \eta_{f,2} -\eta_{b,2} \\
& \stackrel{(b)}{\leq}& n_f I(V;X_f)+ n_b I(W_{1};Y_b) - 3 n_f \alpha + n_f [I(V;Z_f) - I(V;X_f)] \\
&& + n_b [I(W_1;Z_b|W_2)-I(W_1;Y_b|W_2)] - n_{b,2} I(W_2;Y_b) - n_{b,1} I(W_2;Y_b) \\
&=&   n_b I(W_{1};Y_b) - 3 n_f \alpha + {n_f} I(V;Z_f)+  n_b [I(W_1;Z_b|W_2)-I(W_1;Y_b|W_2)] - n_b I(W_2;Y_b) \\
&\stackrel{(c)}{=}&   {n_f} I(V;Z_f) + n_b I(W_1;Z_b|W_2) - 3 n_f \alpha \\
&\stackrel{(d)}{<}&   {n_f} I(V;Z_f) + n_b I(W_1;Z_b) - 9 N \epsilon.
\end{eqnarray*}
Equality (a) follows from (\ref{eta_1-def}) and (\ref{kappa&gamma-def}), inequality (b) follows from the definition of $R_{sk}$ in (\ref{R_s}), equality (c) is due to the Markov chain $W_2 \leftrightarrow W_1 \leftrightarrow Y_b$, and inequality (d) is due to the Markov chain $W_2 \leftrightarrow W_1 \leftrightarrow Z_b$. Since $\gamma-\eta_2$ is sufficiently smaller than ${n_f} I(V;Z_f) +  n_b I(W_1;Z_b)$, from joint-AEP for bipartite sequences \cite[Theorem 4]{Ah10}, for an appropriately chosen partition $\{\mathcal{G}_i\}_{i=1}^{2^\kappa}$, the decoding error probability becomes arbitrarily close to 0, i.e., given $(S,F_2,B_2,Z_f^{n_f},Z^{n_b}_b)$,

\[ \Pr\left((\check{V}^{n_f},\check{W}_1^{n_{b}}) \neq (V^{n_f}, W^{n_b}_1)\right) < 2 \epsilon. \]

Letting $\check{F}=\mathfrak{f}(\check{V}^{n_f})$ and $\check{B},\check{F}=Enc(\check{W}_1^{n_{b}})$, we have
\begin{eqnarray*}
\Pr\left((\check{F},\check{B}) \neq (F, B)\right) < 2 \epsilon.
\end{eqnarray*}
Using Fano's inequality \cite{Ga68} results in
\begin{eqnarray}
H(F,B|S,F,B,Z_f^{n_f},Z_b^{n_b})\leq H(F,B|\check{F},\check{B}) < h(2\epsilon)+2\epsilon \eta, \label{HS-bound4}
\end{eqnarray}
where $h(\epsilon)=-\epsilon\log(\epsilon)-(1-\epsilon)\log(1-\epsilon)$ is the binary entropy function. Applying (\ref{HS-bound2})-(\ref{HS-bound4}) in (\ref{HS-bound1}) gives
\begin{eqnarray*}
H(S|Z_f^{n_f},Z^{n_b}_b) & >& n_f [I(V;X_f)-I(V;Z_f)] + n_b [I(W_{1};Y_b|W_{2})- I(W_{1};Z_b|W_{2})] \\
&& - 14 N \epsilon -h(2\epsilon)-2\epsilon \eta\\
&=& ({n_f}+{n_b})R_{sk}- 14 N \epsilon -h(2\epsilon)-2\epsilon \eta\\
&\geq& H(S) - 14 N \epsilon -h(2\epsilon)+2\epsilon \eta,
\end{eqnarray*}
where the last inequality follows from (\ref{H_S}). This implies that by appropriate selection of $\epsilon$ for an arbitrarily small $\delta$, we will have
\begin{eqnarray*}
\frac{H(S|Z_f^{n_f},Z^{n_b}_b) }{H(S)} > 1-\delta.
\end{eqnarray*}

\subsection{Proof of Proposition \ref{proposition-lowerbound}}\label{sec-sd-lowerbound}
From (\ref{Markov-Chain-1}) and the independence of the two DMCs in the sd-2DMBC setup (see Definitions \ref{definition-stoch-degraded} and \ref{definition-sd-2DMBC}), $V_f\leftrightarrow Y_f \leftrightarrow  X_f \leftrightarrow Z_f$ forms a Markov chain, and so we write (\ref{RA_{sk}1}) and (\ref{RB_{sk}1}) as
\begin{IEEEeqnarray}{lll} \label{RA_{sk}1-G}
R^A_{s1}&=&I(V_f;X_f,Z_f)-I(V_f;Z_f)=I(V_f;X_f|Z_f),\\
R^B_{s1}&=&I(V_b;X_b,Z_b)-I(V_b;Z_b)=I(V_b;X_b|Z_b).
\end{IEEEeqnarray}

From Definition \ref{definition-stoch-degraded} and the second Markov chain in (\ref{Markov-Chain-1}), there exist $\tilde{Y}_b$ and $\tilde{Z}_b$ such that one of the Markov chains
\begin{IEEEeqnarray}{l} \label{SD-deg-chains}
W_{2,b} \leftrightarrow W_{1,b} \leftrightarrow X_b \leftrightarrow \tilde{Y}_b \leftrightarrow \tilde{Z}_b, \mbox{ or} \IEEEyessubnumber \label{obverse-deg}\\
W_{2,b} \leftrightarrow W_{1,b} \leftrightarrow X_b \leftrightarrow \tilde{Z}_b \leftrightarrow \tilde{Y}_b \IEEEyessubnumber \label{reverse-deg}
\end{IEEEeqnarray}
hold, and
\begin{IEEEeqnarray*}{l}
I(X_b;Y_b) = I(X_b;\tilde{Y}_b),~~ I(X_b;Z_b) = I(X_b;\tilde{Z}_b) \\
I(W_{1,b};Y_b|W_{2,b}) = I(W_{1,b};\tilde{Y}_b|W_{2,b}),\\
I(W_{1,b};Z_b|W_{2,b}) = I(W_{1,b};\tilde{Z}_b|W_{2,b}).
\end{IEEEeqnarray*}
Hence, we write (\ref{RA_{sk}2}) as
{\small
\begin{IEEEeqnarray}{l} \label{RA_{sk}2-G}
R^A_{s2}=                     I(W_{1,b};\tilde{Y}_b|W_{2,b})-I(W_{1,b};\tilde{Z}_b|W_{2,b}) \nonumber \\
~        \leq  I(W_{1,b};\tilde{Y}_b|\tilde{Z}_b,W_{2,b}) \stackrel{(a)}{\leq}  I(X_b;\tilde{Y}_b|\tilde{Z}_b) \nonumber\\
~        =                     [I(X_b;\tilde{Y}_b) - I(X_b;\tilde{Z}_b)]_+ = [I(X_b;Y_b) - I(X_b;Z_b)]_+. \tab
\end{IEEEeqnarray}
}
Inequality (a) follows from (\ref{SD-deg-chains}). More precisely, if (\ref{obverse-deg}) holds the inequality is easily satisfied, and if (\ref{reverse-deg}) holds both sides equal zero. It is easy to see that equality in (\ref{RA_{sk}2-G}) holds by choosing $W_{2,b}=1$ and $W_{1,b}$ to be $X_b$ or $1$, in the case of (\ref{obverse-deg}) or (\ref{reverse-deg}), respectively. In analogy to the above, we have
\begin{IEEEeqnarray}{l}
R^B_{s2} \leq [I(X_f;Y_f)-I(X_f;Z_f)]_+,
\end{IEEEeqnarray}
where equality holds for some $W_{2,f}$ and $W_{1,f}$. By replacing $R^A_{s1},R^A_{s2},R^B_{s1},$ and $R^B_{s2}$ in (\ref{L_A}) and (\ref{L_B}) with the above-obtained quantities, (\ref{lower-bound}) is simplified to (\ref{SD-lower-bound}).

\subsection{Proof of Theorem \ref{theorem-SD-2DMBC}} \label{sec-sd-capacity}
\begin{figure} [h]
\centering
  \includegraphics[scale=.4]{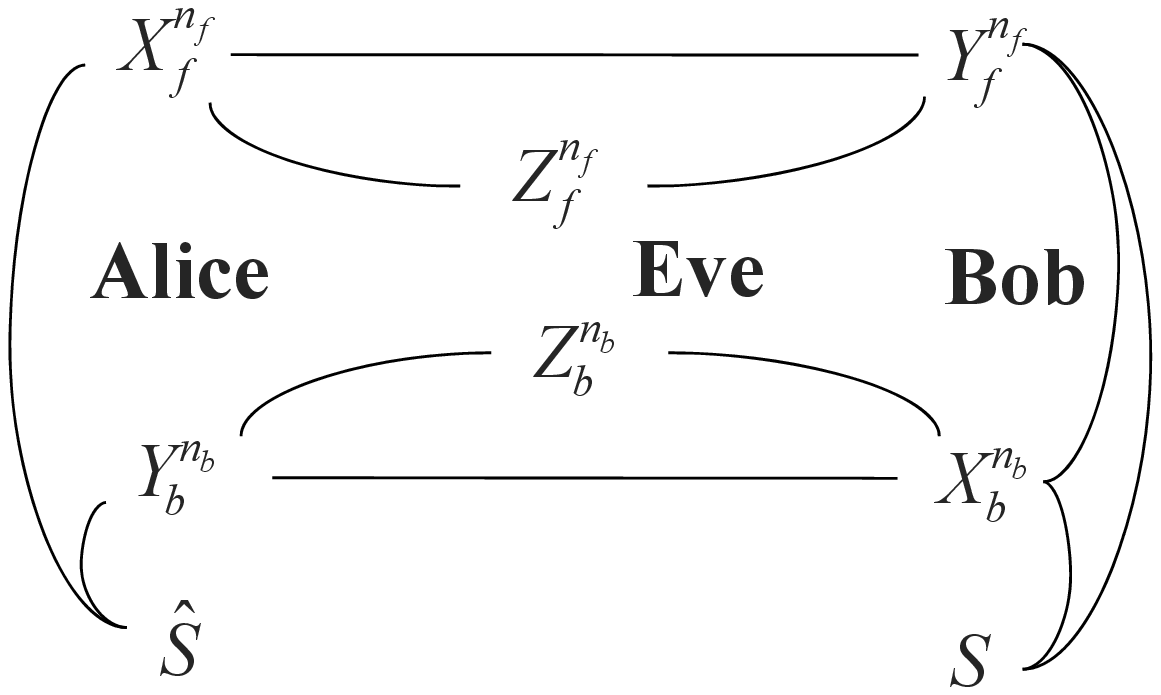}
  \caption{The relations between variables/sequences in two-round SKE when Alice starts the protocol and Bob calculates the key}\label{fig-two-round SKE}
\end{figure}

We let Alice be the party who sends i.i.d. variables. The other case follows by symmetry. We use Lemma \ref{lemma1} to reduce a multi-round SKE protocol to a two-round one, and then give the highest rate that a two-round protocol can achieve.
\begin{lemma}\label{lemma1}
When Alice can only send i.i.d. variables, the secret-key capacity is achieved by a two-round SKE protocol whose initiator is Alice.
\end{lemma}
\begin{IEEEproof}
Let $\Pi$ be a $t$-round SKE protocol that achieves the secret-key capacity under the above condition.\\
\emph{Case 1: Alice sends in odd rounds.} In any (odd) round $r$, Alice's sent sequence $X_f^{:r}$ is independent of her view in round $r-1$, and hence she could compute it in the first communication round. Besides, sending this sequence in the first round does not affect the distribution of Bob's and Eve's received sequences ($Y_f^{:r}$ and $Z_f^{:r}$) since the channels are memoryless. Obviously Bob can compute $X_b^{:r}$ for any even $r$ as before. Hence, we can convert the protocol $\Pi$ into $\Pi'$ in which Alice sends the whole $||_{(odd) r\leq t} \left[X^{n_{f,r}:r}_f \right]$ in the first round such that all the communicated sequences and the final key in $\Pi$ and $\Pi'$ have the same joint probability distribution, i.e., if the same randomness is chosen by Alice, Bob, and the 2DMBC in the execution of $\Pi$ and $\Pi'$, then all the communicated sequences and the final key are identical. Now, Bob can send the whole $||_{(even) r\leq t} \left[X^{n_{b,r}:r}_b \right]$ in the second round without affecting the joint distribution of the sequences. We refer to this last protocol as $\Pi''$ which is a two-round protocol with Alice as the initiator such that the communicated sequences and the key have the same joint distribution as in $\Pi$. Hence $\Pi''$ achieves the secret-key capacity.
\\
\emph{Case 2: Alice sends in even rounds.} Using a similar argument to that of Case 1, we reach a three-round protocol $\Pi''$ with Bob as the initiator: Bob sends $X^{n_{b,1}:1}_b$ in the first round, Alice sends $||_{(even) r\leq t} \left[X^{n_{f,r}:r}_f \right]$ in the second round, and Bob sends $||_{(odd) 3\leq r\leq t} \left[X^{n_{b,r}:r}_b \right]$ in the third round. Since the communicated sequence in the first round is not used to calculate the second round communicated sequences, Bob can send $X^{n_{b,1}:1}_b$ in the third round without affecting the distribution of the sequences in the protocol $\Pi''$. This gives a two-round communication protocol with Alice as the initiator that achieves the capacity.
\end{IEEEproof}
Now, consider a two-round SKE protocol as depicted in Fig. \ref{fig-two-round SKE} in which Alice sends a sequence of i.i.d. variables $X_f^{n_f}$ in the first round. Since the channels are memoryless and independent, Bob and Eve receive sequences of i.i.d. variables $Y_f^{n_f}$ and $Z_f^{n_f}$ and  $Y_f\leftrightarrow X_f \leftrightarrow Z_f$ is a Markov chain. This can be seen as the Discrete Memoryless Multiple Source (DMMS) $(Y_f,X_f,Z_f)$ between Bob, Alice, and Eve, respectively and the DMBC $X_b\rightarrow (Y_b,Z_b)$ from Bob to Alice and Bob. When the DMMS and DMBC satisfy the degradedness condition $Y_f\leftrightarrow X_f \leftrightarrow Z_f$ and $X_b\leftrightarrow Y_b \leftrightarrow Z_b$, \cite{Kh08} proves an upper bound on the secret-key capacity that coincides with the lower bound in (\ref{SD-lower-bound}). However, the proof in \cite{Kh08} can not be directly applied to our problem due to the ``stochastic'' degradedness of the (backward) DMBC. We give the following argument to upper bound the highest achievable rate $R_{sk}$ for an arbitrarily small $\delta>0$ as in (\ref{SKE-Eqs}).

The views of the parties at the end of the second round are $View_A=(X^{n_f}_f,Y^{n_b}_b)$, $View_B=(Y^{n_f}_f,X^{n_b}_b)$, and $View_E=(Z^{n_f}_f,Z^{n_b}_b)$. Using Fano's inequality for (\ref{SKE-rel}), we have
\begin{eqnarray}
H(S|View_A) \leq H(S|\hat{S}) < h(\delta) + \delta H(S) , \label{H(S|S')}
\end{eqnarray}
Furthermore, (\ref{SKE-sec}) gives
\begin{eqnarray}\label{I(S;Z)}
I(S;View_E)=H(S)-H(S|View_E) \leq \delta H(S).
\end{eqnarray}
In the following, we omit the length of the sequences, $X_f^{n_f},Y_f^{n_f},Z_f^{n_f}$ and $X_b^{n_b},Y_b^{n_b},Z_b^{n_b}$ from the superscripts, instead use bold to denote them. $H(S)$ is upper bounded as
\begin{IEEEeqnarray}{rll} \label{H(S)-2}
H(S)    &=&    I(S;View_A) + H(S|View_A) \nonumber \\
        &\stackrel{(a)}{\leq}&  I(S;View_A) - I(S;View_E) + h(\delta) + 2 \delta H(S) \nonumber \\
        &\leq&  I(S;View_A | View_E) + h(\delta) + 2 \delta H(S) \label{H(S)-1}\nonumber \\
\Rightarrow (1-2\delta)H(S)-h(\delta) &\leq&  I(S;View_A) - I(S;View_E) \nonumber\\
&=&  I(S;\mathbf{Y}_b) + I(S;\mathbf{X}_f|\mathbf{Y}_b) - I(S;\mathbf{Z}_f,\mathbf{Z}_b) \nonumber \\
&=&  I(S;\mathbf{Y}_b)+I(S;\mathbf{X}_f,\mathbf{Z}_f|\mathbf{Y}_b)-I(S;\mathbf{Z}_f,\mathbf{Z}_b) \nonumber\\
&=&  I(S;\mathbf{Y}_b)+I(S;\mathbf{Z}_f|\mathbf{Y}_b)+I(S;\mathbf{X}_f|\mathbf{Z}_f,\mathbf{Y}_b)-I(S;\mathbf{Z}_f,\mathbf{Z}_b) \nonumber\\
&=&  [I(S;\mathbf{Z}_f,\mathbf{Y}_b)-I(S;\mathbf{Z}_f,\mathbf{Z}_b)]+[I(S;\mathbf{X}_f|\mathbf{Z}_f,\mathbf{Y}_b)],
\end{IEEEeqnarray}
where inequality (a) follows from (\ref{H(S|S')}) and (\ref{I(S;Z)}). We separately discuss the two terms in (\ref{H(S)-2}). Note that $(S,\mathbf{Z}_f) \leftrightarrow \mathbf{X}_b \leftrightarrow (\mathbf{Y}_b,\mathbf{Z}_b)$ is a Markov chain. If the backward DMBC is stochastically degraded in favor of $Z_b$, the first term is at most zero; otherwise, letting $X_b \leftrightarrow \tilde{Y}_b \leftrightarrow \tilde{Z}_b$ (see Definition \ref{definition-stoch-degraded}), we have
\begin{IEEEeqnarray}{lll}\label{I(S:...)-1}
I(S;\mathbf{Z}_f,\mathbf{Y}_b)-I(S;\mathbf{Z}_f,\mathbf{Z}_b) &=& I(S;\mathbf{Z}_f,\tilde{\mathbf{Y}}_b)-I(S;\mathbf{Z}_f,\tilde{\mathbf{Z}}_b) \nonumber \\
&=&  I(S;\mathbf{Z}_f,\tilde{\mathbf{Y}}_b,\tilde{\mathbf{Z}}_b)-I(S;\mathbf{Z}_f,\tilde{\mathbf{Z}}_b) I(S;\tilde{\mathbf{Y}}_b|\mathbf{Z}_f,\tilde{\mathbf{Z}}_b) \nonumber \\
&\leq& I(S,\mathbf{Z}_f;\tilde{\mathbf{Y}}_b|\tilde{\mathbf{Z}}_b) = I(S,\mathbf{Z}_f;\tilde{\mathbf{Y}}_b) - I(S,\mathbf{Z}_f;\tilde{\mathbf{Z}}_b) \nonumber \\
&=&  I(S,\mathbf{Z}_f;\mathbf{Y}_b) - I(S,\mathbf{Z}_f;\mathbf{Z}_b) \stackrel{(a)}{\leq}  n_b [I(W_b;Y_b) - I(W_b;Z_b)] ~~ \nonumber \\
&\stackrel{(b)}{\leq}& n_b [I(X_b;Y_b) - I(X_b;Z_b)]_+.
\end{IEEEeqnarray}
Inequality (a) follows from the results of message transmission over single DMBCs (e.g., \cite[Section V]{Cs78}), where the conditional distribution $P_{Y_b,Z_b|X_b}$ corresponds to the backward DMBC and $W_b$ is an RV that satisfies the Markov chain $W_b \leftrightarrow X_b \leftrightarrow (Y_b,Z_b)$. Inequality (b) is due to the degradedness of the backward DMBC.
Letting $J$ be an independent random variable uniformly distributed over $\{1,2,\dots, n_f\}$, we write the second term in (\ref{H(S)-2}) as
\begin{IEEEeqnarray}{lll}
I(S;\mathbf{X}_f|\mathbf{Z}_f,\mathbf{Y}_b) &\leq& I(S, \mathbf{Y}_b;\mathbf{X}_f|\mathbf{Z}_f) \nonumber \\
&\stackrel{(a)}{=}& I(S, \mathbf{Y}_b;\mathbf{X}_f)-I(S, \mathbf{Y}_b;\mathbf{Z}_f) \nonumber\\
&\stackrel{(b)}{=}& \sum_{i=1}^{n_f} I(S, \mathbf{Y}_b;X_{f,i}|Z_{f,i+1}^{n_f},X^{i-1}_{f}) -I(S, \mathbf{Y}_b;Z_{f,i}|Z_{f,i+1}^{n_f},Z^{i-1}_{f}) \nonumber\\
&\stackrel{(c)}{=}& \sum_{i=1}^{n_f} I(S, \mathbf{Y}_b;X_{f,i}|Z_{f,i},Z_{f,i+1}^{n_f},X^{i-1}_{f}) \nonumber\\
&=& n_f I(S, \mathbf{Y}_b;X_{f,J}|Z_{f,J},Z_{f,J+1}^{n_f},X^{J-1}_{f},J) \nonumber \\
&\leq& n_f I(S, \mathbf{Y}_b, Z_{f,J+1}^{n_f},X^{J-1}_{f},J;X_{f,J}|Z_{f,J}). \tab
\end{IEEEeqnarray}
Equality (a) is due to the Makov chain $\mathbf{Z}_f\leftrightarrow \mathbf{X}_f \leftrightarrow (S,\mathbf{Y}_b)$, equality (b) follows from the chain rule for difference between mutual information (see e.g., \cite[Section V]{Cs78}), and equality (c) is due to the Markov chain $Z_{f,i}\leftrightarrow X_{f,i} \leftrightarrow (S,\mathbf{Y}_b)$.

Now, letting $V_f=(S, \mathbf{Y}_b, Z_{f,J+1}^{n_f},X^{J-1}_{f},J)$, $X_f=X_{f,J}$, $Y_f=Y_{f,J}$ and $Z_f=Z_{f,J}$, the conditional distribution $P_{Y_f.Z_f|X_f}$ corresponds to the forward DMBC, the Markov chain $Z_f \leftrightarrow X_f\leftrightarrow Y_f \leftrightarrow V_f$ is satisfied, and we have
\begin{IEEEeqnarray}{l}\label{I(S:...)-2}
I(S;\mathbf{X}_f|\mathbf{Z}_f,\mathbf{Y}_b)\leq n_f I(V_f;X_f|Z_f).
\end{IEEEeqnarray}
Using the quantities of (\ref{I(S:...)-1}) and (\ref{I(S:...)-2}) in the calculation of (\ref{H(S)-2}), $H(S)$ is upper bounded as
\begin{IEEEeqnarray}{lll}
H(S) &\leq& \frac{n_f I(V_f;X_f|Z_f)+ n_b [I(X_b;Y_b)-I(X_b;Z_b)]_++h(\delta)}{(1-2\delta)}  \nonumber \\
&=& n_f I(V_f;X_f|Z_f)+ n_b [I(X_b;Y_b)-I(X_b;Z_b)]_+,\label{H(S)-final}
\end{IEEEeqnarray}
where the last equality holds since $\delta$ is arbitrarily small. This together with (\ref{SKE-rand}) proves
the argument in (\ref{L'_A}), and the condition in (\ref{L'_A}) is proven as follows.
\begin{IEEEeqnarray}{lll} \label{nI(xb;yb)}
n_b I(X_b;Y_b) &\geq& I(\mathbf{X}_b;\mathbf{Y}_b) \stackrel{(a)}{\geq} I(\mathbf{Y}_f;\mathbf{Y}_b) \nonumber \\
 &=& I(\mathbf{Y}_b,S;\mathbf{Y}_f)-I(S;\mathbf{Y}_f|\mathbf{Y}_b) \geq I(\mathbf{Y}_b,S;\mathbf{Y}_f)-H(S|\mathbf{Y}_b) \nonumber \\
 &=& I(\mathbf{Y}_b,S;\mathbf{Y}_f)-H(S|\mathbf{Y}_b,\mathbf{X}_f)-I(S;\mathbf{X}_f|\mathbf{Y}_b) \nonumber \\
 &\stackrel{(b)}{\geq}& I(\mathbf{Y}_b,S;\mathbf{Y}_f)  - h(\delta) - \delta H(S) - I(S;\mathbf{X}_f|\mathbf{Y}_b) \nonumber \\
 &\stackrel{(c)}{\geq}& I(\mathbf{Y}_b,S;\mathbf{Y}_f)- I(\mathbf{Y}_b,S;\mathbf{X}_f) \nonumber \\
&\stackrel{(d)}{=}& \sum_{i=1}^{n_f} I(\mathbf{Y}_b,S,X^{i-1}_f,Y^{n_f}_{f,i+1};Y_{f,i}) - I(\mathbf{Y}_b,S,X^{i-1}_f,Y^{n_f}_{f,i+1};X_{f,i}) \nonumber \\
&\stackrel{(e)}{=}& \sum_{i=1}^{n_f} I(\mathbf{Y}_b,S,X^{i-1}_f,Y^{n_f}_{f,i+1};Y_{f,i}|X_{f,i}) \nonumber \\
&\stackrel{(f)}{\geq}& \sum_{i=1}^{n_f} I(\mathbf{Y}_b,S,X^{i-1}_f,Z^{n_f}_{f,i+1};Y_{f,i}|X_{f,i}) \nonumber \\
&=& n_f I(\mathbf{Y}_b,S,X^{J-1}_f,Z^{n_f}_{f,J+1};Y_{f,J}|X_{f,J},J) = n_f I(V_f;Y_{f}|X_{f})- n_f I(J;Y_{f}|X_{f}) \nonumber \\
&\stackrel{(g)}{=}& n_f I(V_f;Y_{f}|X_{f}).
\end{IEEEeqnarray}
Inequality (a) is due to the Markov chain $\mathbf{Y}_f \leftrightarrow \mathbf{X}_b\leftrightarrow \mathbf{Y}_b$; inequality (b) follows from (\ref{H(S|S')}); inequality (c) holds since $\delta$ is arbitrarily small and so $h(\delta)+\delta H(S)$ is negligible compared to the other quantities; equality (d) follows from the chain rule for difference between mutual information; equality (e) is due to the Markov chain $X_{f,i} \leftrightarrow Y_{f,i}\leftrightarrow (\mathbf{Y}_b,S,X^{i-1}_f,Y^{n_f}_{f,i+1})$; inequality (f) is due to the Markov chain $Z^{n_f}_{f,i+1} \leftrightarrow Y^{n_f}_{f,i+1}\leftrightarrow Y_{f,i}$, and equality (g) holds since $Y_{f,J}$ is (i.i.d.) independent of $J$.

One can prove (\ref{L'_B}) by symmetry. This implies that, under the condition of this theorem, equality in (\ref{SD-lower-bound}) holds.

\section{Conclusion}\label{sec_conclusion}
We extended the results of SKE in the 2DMBC setup in the following two cases. When both DMBCs have secrecy potential, we proposed the interactive channel coding (ICC) protocol and proved that it achieves the lower bound. When both DMBCs are stochastically degraded with independent channels (so called sd-2DMBC), we provided a simplified expression for the lower bound, and proved that this lower bound is tight under the condition that one of the parties sends only i.i.d variables. Obtaining a single-letter characterization or even a tighter upper bound for the secret-key capacity in the sd-2DMBC setup remains as future work.

\end{document}